\documentclass[conference]{IEEEtran}
\IEEEoverridecommandlockouts
\usepackage{amsmath}
\usepackage[caption=false]{subfig}
\usepackage{tabularx,graphicx}
\usepackage{hyperref}
\usepackage[numbers,sort&compress]{natbib}

\begin{document}

\title{The Evaluation of Rating Systems\\ in Team-based Battle Royale Games}


\author{\IEEEauthorblockN{Arman Dehpanah}
\IEEEauthorblockA{\textit{School of Computing} \\
\textit{DePaul University}\\
Chicago, USA \\
\small{adehpana@depaul.edu}}
\and
\IEEEauthorblockN{Muheeb Faizan Ghori}
\IEEEauthorblockA{\textit{School of Computing} \\
\textit{DePaul University}\\
Chicago, USA \\
\small{mghori2@depaul.edu}}
\and
\IEEEauthorblockN{Jonathan Gemmell}
\IEEEauthorblockA{\textit{School of Computing} \\
\textit{DePaul University}\\
Chicago, USA \\
\small{jgemmell@cdm.depaul.edu}}
\and
\IEEEauthorblockN{Bamshad Mobasher}
\IEEEauthorblockA{\textit{School of Computing} \\
\textit{DePaul University}\\
Chicago, USA \\
\small{mobasher@cs.depaul.edu}}
}
 
\maketitle

\begin{abstract}

Online competitive games have become a mainstream entertainment platform.
To create a fair and exciting experience, these games use rating systems to match players with similar skills.
While there has been an increasing amount of research on improving the performance of these systems, less attention has been paid to how their performance is evaluated.
In this paper, we explore the utility of several metrics for evaluating three popular rating systems on a real-world dataset of over 25,000 team battle royale matches.
Our results suggest considerable differences in their evaluation patterns.
Some metrics were highly impacted by the inclusion of new players.
Many could not capture the real differences between certain groups of players.
Among all metrics studied, normalized discounted cumulative gain (NDCG) demonstrated more reliable performance and more flexibility.
It alleviated most of the challenges faced by the other metrics while adding the freedom to adjust the focus of the evaluations on different groups of players.

\end{abstract}


\section{Introduction}

Online competitive games have been continuously growing over the past years.
Today, games such as League of Legends or PlayerUnknown's Battlegrounds attract hundreds of millions of players from all around the world.
Player-versus-Player (PvP) is one of the most popular game-play modes of such games where players compete against human opponents.
To make sure the competition is fair and engaging, these games match players with similar skills.
Many PvP games use rating systems for this purpose.

Rating systems leverage numerical representations to describe a player's skills.
These representations are updated after each match based on the outcomes and converge to the true skill level of the players as the system observes more games from them.
Rating systems use skill ratings to predict rank.
Numerous efforts have been made to achieve more accurate skill estimations for improving rank prediction.
However, less attention has been paid to the evaluation of predicted ranks.

Previous efforts leveraged a variety of metrics to evaluate predicted ranks.
While these metrics are able to make a distinction between the predictive performance of rating systems, they often provide global evaluations without regard to what the rankings actually represent.
Some metrics may not consider the ordinal nature of ranks while others may treat high-ranked players the same as low-ranked players. 

In this paper, we consider traditional metrics such as accuracy, mean absolute error (MAE), and Kendall's rank correlation coefficient.
We also include mean reciprocal rank (MRR), average precision (AP), and normalized discounted cumulative gain (NDCG)-- metrics that are commonly used for evaluating query-response systems in the field of information retrieval.
We explore how well these metrics capture different aspects of the evaluation of rank prediction.
To perform this analysis, we consider three popular rating systems; Elo, Glicko, and TrueSkill.
We limit our experimentation to team-based battle royale games-- a popular game-play mode where several teams simultaneously compete against one another in the same match and the last team standing wins the game.
Our real-world dataset includes over 25,000 team-based battle royale matches and over 825,000 unique players from PlayerUnknown's Battlegrounds.

Our experiments show that metrics adapted from the field of information retrieval can better evaluate the performance of rating systems in team-based battle royale games.
In particular, NDCG more accurately demonstrates both the predictive power and predictive behavior of these systems.
NDCG evaluations are more reliable since it better captures the difference between players and teams by applying a position-based weight.
These weights enable NDCG to place more emphasis on achieving correct predictions for higher ranks, an important utility as these ranks often correspond to top-tier players who are more engaged with the game.



\section{Related Work}
\label{sec:related}
Reports suggest that the video game industry is now more profitable than the music and movie industries combined~\cite{website2}.
A large share of this market is accounted for by online competitive games.

These games often leverage rating systems to rate players' skills.
For example, Elo~\cite{elo1978rating} uses a single arbitrary number for representing the player's skills while Glicko~\cite{glickman1995glicko} and TrueSkill~\cite{herbrich2007trueskill} assume the players' skills follow a Gaussian distribution with distinct means and standard deviations.
On the other hand, some systems leverage historical in-game statistics to represent players' skills.
Depending on the game's genre, these statistics may include \textit{kill to death ratio}, \textit{distance walked}, \textit{role}, or \textit{gold owned} among others~\cite{nikolakaki2020competitive, suznjevic2015application, hodge2019win}.

One of the main applications of rating systems in online competitive games is to match players based on their skill levels to create balanced matches.
Given a set of players and their skill ratings, these systems test different match-ups, predict potential outcomes, and launch the matches where each side has an equal chance of winning.
Those predictions can then be evaluated to assess the performance of rating systems. 

Numerous efforts have been made to evaluate the predictive power of different approaches of skill rating in head-to-head games.
In these games, two players or teams compete against one another and the winner is the side that gains the highest number of points after a certain condition is met, e.g., reaching a certain amount of time or a certain number of rounds.
Since the outcome of these games is either win or loss (or draw when possible), the majority of works focused on how accurately the systems classify players and teams into their observed outcome.  
Accuracy~\cite{dehpanah2021evaluating,bisberg2019scope, gong2020optmatch, guo2012score}, F1 score~\cite{nikolakaki2020competitive, aung2018predicting}, and log likelihood~\cite{chen2016predicting, stanescu2011rating} are among the metrics commonly used for evaluating these works.
In addition, some works focused on how closely the systems predict the ratings and their associated ranks compared to the observed outcomes.
Mean absolute error~\cite{gong2020optmatch, guo2012score, yu2018moba}, mean squared error~\cite{lasek2013predictive, yu2018moba}, and root mean square error~\cite{gong2020optmatch} are among the metrics commonly used for evaluating rank prediction in this group of research.

Efforts have also been made to evaluate rank predictions in free-for-all games where the competition involves more than two sides.
These games can be divided into two groups: deathmatch and battle royale.
In deathmatch, players eliminate each other and the winner is the player or team with the highest score at the end of the match.
In battle royale matches, several players or teams compete against one another at the same time.
Players start eliminating each other and the winner is the last player or team standing.
Since the outcome of free-for-all games is a ranked list of players or teams, to evaluate rank predictions, most of the previous works focused on the ordinal association between the predicted and observed ranks.
Rank correlation coefficients and their derivatives~\cite{shafiee2012study, herbrich2007trueskill, buckley2015rapid, weng2011bayesian} are among the metrics commonly used for this purpose.
A recent study compared several metrics based on how they explain the predictive power of rating systems~\cite{dehpanah2020evaluation}. 
This study focused on solo battle royale matches where there are no teams, and players compete against each other as singletons.
The outcome of these matches is directly determined by the performance of each individual player.
However, in a team-based battle royale match, where multiple teams compete against each other, the outcome is highly influenced by the quality of interactions between team members.
Therefore, estimating the true skill level of players and predicting their rank in these games become more labyrinthine compared to solo games.
Nevertheless, the utility of different metrics for evaluating rank prediction in team-based battle royale is yet to be explored.

Our work differs from previous efforts.
We focus on team battle royale, a game-play mode that has not received enough attention in the research literature despite its increasing popularity among players.
We extend Elo and Glicko to team-based battle royale games.
Using a large real-world dataset, we investigate the explanatory power of six metrics, including traditional metrics and those adapted from the field of information retrieval.
We analyze how these metrics operate under the influence of new players.
We also consider how these metrics describe different groups of players such as top-tier and the most frequent players.


\section{Rating Systems}
\label{sec:systems}

Rating systems have become an integral part of online competitive games.
These systems leverage a numeric value, referred to as skill rating and denoted by $\mu$, to represent the skill level of players.
The core function of such systems is to predict ranks by calculating the probability of winning for each competing player in a match based on their skill ratings.
The ratings are then updated after the match by comparing the predicted ranks, $R^{pred}$, with the observed ranks, $R^{obs}$.

The updating method varies depending on the rating system.
In the remainder of this section, we describe three common algorithms: Elo, Glicko, and TrueSkill.


\subsection{Elo}
\label{sec:elo}

Elo is a popular rating system that works based on statistical estimation~\cite{elo1978rating}.
Based on the theory behind Elo, the skill level of players can be inferred from the outcome of the games they play.
It follows a self-correcting mechanism in the sense that the early estimations will be modified over time by observing more outcomes.
This way, playing more games will result in more accurate skill estimations.

Elo assumes that players' skills follow a Gaussian distribution with the mean $\mu$, referred to as skill rating.
For simplicity, Elo considers a fixed standard deviation for all players.
The outcome of a match between two players can be predicted using the difference in their skill ratings.
The ratings are updated after each match with the winner capturing points from the loser's rating.
The amount of the points transferred depends on the predicted outcome of the match.







Elo was originally designed for head-to-head matches between two players.
It was later extended to solo battle royale games where many players compete against one another as singletons at the same time in the same match~\cite{dehpanah2020evaluation}.

We further extend Elo to team-based battle royale matches.
To this end, we follow the assumption introduced by TrueSkill~\cite{herbrich2007trueskill} that the overall rating of a team is equal to the sum of the ratings of its members.

Assume \textit{N} teams $t_1$, $t_2$, ..., $t_n$ compete against each other in a field \textit{F} denoting a battle royale match.
If team $t_i$ consists of players $p_{1}$, $p_{2}$, ..., $p_{n}$ with ratings of $\mu_{1}$, $\mu_{2}$, ..., $\mu_{n}$, its overall rating, $\mu_{t_i}$, can be calculated as:

\small
\begin{equation*}
    \mu_{t_i} = \sum^{n}_{j=1}{\mu_j}
\end{equation*}
\normalsize

\noindent where \textit{n} is the total number of players in the team.
We then calculate a contribution weight for each team member denoting the degree to which they contribute to the team's overall rating.
For player $p_j$, this weight is calculated as:

\small
\begin{equation*}
    w_{p_j} = \frac{\mu_j}{\mu_{t_i}}
\end{equation*}
\normalsize

We can consider the match as a set of head-to-head matches.
We calculate the probability of winning for each team by summing the probability of winning values in all their pairwise matches versus the other teams.
The overall probability of winning for team $t_i$ can be calculated as:

\small
\begin{equation*}
    Pr(t_i\:wins, F) = \frac{\sum\limits_{1 \leq j \leq N, i \neq j }  \big({1 + e^{\frac{(\mu_{t_j} - \mu_{t_i})}{D}}}\big)^{-1}}{{\binom{N}{2}}} 
\end{equation*}
\normalsize

\noindent where \textit{D} determines the influence of the difference between ratings, and $\binom{N}{2}$, the total number of pairwise comparisons, is used to normalize the probability values to sum up to 1.

Elo follows a zero-sum logic meaning that the total number of awarded points is equal to the total number of deducted points.
Therefore, the observed outcomes should be normalized to sum up to 1.
For team $t_i$, the observed rank $R^{obs}_{t_i}$ is converted into a normalized result, $R^{'}_{t_i}$, calculated as:

\small
\begin{equation*}
\label{eqn:R}
    R^{'}_{t_i} = \frac{N - R^{obs}_{t_i}}{\binom{N}{2}}
\end{equation*}
\normalsize

The rating of team $t_i$ can then be updated as:

\small
\begin{equation*}
    \mu_{t_i}^\prime = \mu_{t_i} + K[R^{'}_{t_i} - Pr(t_i\:wins, F)]
\end{equation*}
\normalsize

Team members receive different shares of the change in the overall team's rating based on their contribution weights.
To calculate the updated rating for each team member, we multiply the change in the team's rating by their corresponding contribution weight and add the product to their previous ratings.
The rating of player $p_j$ is updated as:

\small
\begin{equation*}
\label{eqn:elo_update2}
    \mu_{j}^\prime = \mu_{j} + w_{p_j}(\mu_{t_i}^\prime - \mu_{t_i})
\end{equation*}
\normalsize

Elo is popular for delivering fairly accurate rankings despite its simplicity and naive assumptions.
However, considering a fixed variance for skill ratings could result in reliability issues.
Such issues were addressed by ensuing research works that led to the introduction of several new rating systems.


\subsection{Glicko}
\label{sec:glicko}

Glicko rating system is similar to Elo in that it assumes players' skills follow a Gaussian distribution.
However, Glicko extended upon Elo by introducing rating deviation, $\sigma$, a dynamic parameter determining the accuracy of ratings.

In Glicko, players are represented by two values, mean of the distribution $\mu$ signifying their skill ratings, and deviation $\sigma$ indicating the uncertainty about their ratings.
The $\sigma$ values decrease over time as players play more games.
Both $\sigma$ and $\mu$ values are updated after each match by comparing predicted ranks $R^{pred}$ with observed ranks $R^{obs}$.

Similar to Elo, Glicko was originally designed for head-to-head matches between two players and was later extended to solo battle royale matches~\cite{dehpanah2020evaluation}.
We extend Glicko to team-based battle royale games as we did with Elo.
We consider the rating and rating deviation of a team as the sum of ratings and rating deviations of its members.
We also consider each battle royale match with \textit{N} teams as $\binom{N}{2}$ separate matches between each pair of teams.

Assume \textit{N} teams $t_1$, $t_2$, ..., $t_n$ compete against each other in a field \textit{F} denoting a battle royale match.
If team $t_i$ consists of players $p_{1}$, $p_{2}$, ..., $p_{n}$ with ratings of $\mu_{1}$, $\mu_{2}$, ..., $\mu_{n}$, and rating deviations of $\sigma_{1}$, $\sigma_{2}$, ..., $\sigma_{n}$, its overall rating $\mu_{t_i}$ and rating deviation $\sigma_{t_i}$, can be calculated as:

\small
\begin{equation*}
    \mu_{t_i} = \sum^{n}_{j=1}{\mu_{j}}
,\:\:\:\:\:\:
    \sigma_{t_i} = \sum^{n}_{j=1}{\sigma_{j}}
\end{equation*}
\normalsize

\noindent where \textit{n} is the total number of players in the team.

As each team member contributes to both the rating and rating deviation of the team, separate contribution weights can be calculated for the rating and rating deviation of each member.
For player $p_j$ these weights are calculated as:

\small
\begin{equation*}
    w^{\mu}_{p_j} = \frac{\mu_{j}}{\mu_{t_i}}
,\:\:\:\:\:\:
    w^{\sigma}_{p_j} = \frac{\sigma_{j}}{\sigma_{t_i}}
\end{equation*}
\normalsize

We calculate the probability of winning for each team by summing the probability of winning values in all their pairwise matches versus the other teams.
The overall probability of winning for team $t_i$ can be calculated as:

\small
\begin{equation*}
    Pr(t_i\:wins, F) = \frac{\sum\limits_{1 \leq j \leq N, i \neq j } \big({1 + 10^{\frac{-g(\sqrt{\sigma_{t_i}^2 + \sigma_{t_j}^2})(\mu_{t_i} - \mu_{t_j})}{400}}}\big)^{-1}}{\binom{N}{2}}
\end{equation*}
\normalsize

\noindent where \textit{g} is a weighting function based on deviations, and $\binom{N}{2}$ is used to normalize the probability values to sum up to 1.
Similar to Elo, Glicko follows a zero-sum logic meaning the number of points added to winning teams is equal to the number of points deducted from losing teams.
Therefore, we normalize the observed outcomes to sum up to 1 as we did for Elo.
The rating and deviation of team $t_i$ are updated as:

\small
\begin{gather}
    \nonumber \mu_{t_i}^{\prime} = \mu_{t_i} + \frac{0.0057565}{\frac{1}{\sigma_{t_i}^2} + \frac{1}{d^2}} \big[ g(\sigma_{t_j})(R^{'}_{t_i} -  Pr(t_i\:wins, F))\big] \\
    \nonumber \sigma_{t_i}^{\prime} = \sqrt{\big(\frac{1}{\sigma_{t_i}^2} + \frac{1}{d^2}\big)^{-1}}
\end{gather}
\normalsize

\noindent where $d^2$ is a Hessian function of the log marginal likelihood.



Finally, we multiply the change in the team's rating and rating deviation by the corresponding contribution weights of its members and add the products to their previous values to calculate the updated values for each team member.
The rating and rating deviation of player $p_j$ are updated as:

\small
\begin{equation*}
    \mu_{j}^\prime = \mu_{j} + w^{\mu}_{p_j}(\mu_{t_i}^\prime - \mu_{t_i})
,\:\:\:\:\:\:
    \sigma_{j}^\prime = \sigma_{j} + w^{\sigma}_{p_j}(\sigma_{t_i}^\prime - \sigma_{t_i})
\end{equation*}
\normalsize

Similar to Elo, Glicko was originally designed for head-to-head matches with two players.
TrueSkill extended both of these systems and accepts any type of match-up with any number of players.


\subsection{TrueSkill}
\label{sec:trueskill}
TrueSkill was introduced by Microsoft Research to address the limitations of traditional rating systems.
It leverages Bayesian inference, factor graphs, and expectation propagation algorithm to estimate the skill level of players.

Similar to Glicko, TrueSkill assumes that the players' skills follow a Gaussian distribution with mean of $\mu$ denoting the skill rating and rating deviation of $\sigma$ representing the uncertainty of the system about the rating.
These values are updated after each match by comparing predicted ranks with observed ranks.
The updating method of TrueSkill depends on whether a draw is possible.
for a non-draw case, if $\mu_i$, $\mu_j$, $\sigma_i$, and $\sigma_j$ represent skill ratings and deviations of players $p_i$ and $p_j$, assuming player $p_i$ wins the match against player $p_j$, her skill rating is updated by:

\small
\begin{equation*}
    \mu_i^{\prime} = \mu_i + \frac{\sigma^{2}_i}{c}\big[ \frac{N(\frac{t}{c})}{\Phi(\frac{t}{c})}\big]
\end{equation*}
\normalsize

\noindent where $t = \mu_i - \mu_j$ and $c = \sqrt{2\beta^2 + \sigma_i^2 + \sigma_j^2}$. 
$N$ and $\Phi$ denote the probability density and cumulative distribution functions of a standard normal distribution.
The parameter $\beta$ is the scaling factor determining the magnitude of changes to ratings.
Skill deviations are updated by:

\small
\begin{equation*}
    \sigma^{\prime} =  \sigma - \sigma \big(\frac{\sigma^2}{c^2}  \big[\frac{N(\frac{t}{c})}{\Phi(\frac{t}{c})}\big]    \big[\frac{N(\frac{t}{c})}{\Phi(\frac{t}{c})} + t\big] \big)
\end{equation*}
\normalsize

TrueSkill only focuses on the outcome of the match and does not consider the interaction between players.
Microsoft Research later introduced TrueSkill 2~\cite{minka2018trueskill} to address this issue.


\subsection{PreviousRank}

Elo, Glicko, and TrueSkill are parametric approaches that build a model for players by estimating their relative skill level.
To provide a non-parametric approach, we introduce PreviousRank as our naive baseline.
PreviousRank assumes that a player's predicted rank is equal to their observed rank in their previous match.
If a player is new to the system, we assume that their PreviousRank is equal to $\frac{N}{2}$ where \textit{N} is the number of players competing in the match.
To calculate a team's PreviousRank, we simply add up the PreviousRank of each member.
The team with the lowest PreviousRank is predicted to win the match.








\section{Metrics}
\label{sec: metrics}

In this section, we present our metrics for evaluating the predictive power of rating systems in team-based battle royale.


\subsection{Accuracy}

Accuracy is a common metric in classification tasks for evaluating nominal or categorical outcomes.
In a head-to-head match between two players, there are only two outcomes, win or loss (three if a draw is possible).
We can consider each outcome as a label and use accuracy to evaluate the predicted ranks against observed ranks.
The same holds true for head-to-head matches between two teams.

Battle royale matches consist of many teams.
We can consider the problem of ranking in battle royale matches as a multi-label classification with each team's rank being a unique label.
Similar to head-to-head matches, we can use accuracy for evaluating rank prediction in battle royale matches.
As such, accuracy is calculated as the ratio of correctly classified ranks to the total number of teams.

Using accuracy to evaluate rank prediction poses some limitations as it treats all ranks as labels. 
If predicted and observed ranks are the same, that counts as a hit.
However, if a player was predicted to achieve rank 5 and earned rank 6 or 96, it is a miss even though these two scenarios differ greatly.

In addition, since accuracy treats ranks as labels, it does not distinguish between errors in higher ranks and those in lower ranks.
For example, predicting rank 5 for an observed rank of 1 and predicting rank 95 for an observed rank of 90 both are considered a miss without regard to their rank positions.


\subsection{Mean Absolute Error}

Mean absolute error (MAE) is a common metric for evaluating continuous prediction outcomes. 
It calculates the similarity between two sets of values by performing a pairwise comparison of elements and averaging the absolute errors of each pair.
For a battle royale match between \textit{N} teams, MAE can be calculated as:

\small
\begin{equation*}
    MAE = \frac{1}{N} \sum_{i=1}^{N}|R^{pred}_i - R^{obs}_i|    
\end{equation*}
\normalsize

\noindent where $R^{pred}_i$ and $R^{obs}_i$ show predicted and observed ranks for team $t_i$.
Higher MAE values suggest higher dissimilarities between the two rankings.

Unlike accuracy, MAE does not evaluate the predictions on a hit or miss basis.
Instead, it takes into account the difference between values.
As such, predicting rank 96 for an observed rank of 5 imposes much higher errors compared to a predicted rank of 6.
However, similar to accuracy, MAE does not consider rank positions and thus, treats the errors in higher ranks the same as those in lower ranks.
This way, MAE is unable to capture the differences between certain groups of players.
For example, MAE considers the difference between rank 1 and rank 6 (often achieved by top-tier players) to be the same as the difference between rank 90 and 95 (often belonging to low-tier players).


\subsection{Kendall's Rank Correlation Coefficient}

Kendall's rank correlation coefficient, also referred to as Kendall's tau $\tau$, measures the ordinal association between two sets of values considering the number of concordant and discordant pairs of observations.

Assume two rankings of $R^{pred}$ and $R^{obs}$ as the predicted rank and observed rank of teams in a battle royale match.
For two teams, $t_i$ and $t_j$, any pair of observations $(R^{pred}_i, R^{obs}_i)$ and $(R^{pred}_j, R^{obs}_j)$ are concordant if $R^{pred}_i > R^{pred}_j$ and $R^{obs}_i > R^{obs}_j$, or $R^{pred}_i < R^{pred}_j$ and $R^{obs}_i < R^{obs}_j$.
Otherwise, they are considered discordant.

For a battle royale match, Kendall's tau can be calculated as the difference between the number of concordant and discordant pairs of predicted and observed ranks normalized by the total number of pairwise combinations:

\small
\begin{equation*}
    \tau = \frac{n_c - n_d}   {\binom{N}{2}}
\end{equation*}
\normalsize

\noindent where $n_c$ is the number of concordant pairs, $n_d$ is the number of discordant pairs, and \textit{N} is the total number of competing teams.
Kendall's tau is equal to 1 if the orders of predicted and observed ranks completely agree and -1 when the orders completely disagree.
A tau of zero shows that there is no correlation between the two rankings.

While accuracy and MAE look at the deviation between predicted and observed ranks, Kendall's tau considers the pairwise agreement between two rankings.
For example, if two players were predicted to have ranks 5 and 6 and achieved ranks 30 and 50, tau considers this a concordant pair without regard to the deviations in predicted versus observed ranks.
However, like accuracy and MAE, it does not distinguish between higher rank and lower rank errors.


\subsection{Mean Reciprocal Rank}

Mean reciprocal rank (MRR) is a common metric in the field of information retrieval for evaluating the performance of a query-response system that returns a ranked list of responses for a query.
To use MRR for evaluating the performance of rating systems in team battle royale matches, we utilize an extended version of MRR introduced in~\cite{dehpanah2020evaluation}.
In a battle royale match with \textit{N} teams, MRR can be calculated as:

\small
\begin{equation*}
    MRR = \frac{1}{N}\sum_{i=1}^{N}\frac{1}{1 + error_i}
\end{equation*}
\normalsize

\noindent where $error_i$ is the absolute difference between the predicted rank $R^{pred}$ and observed rank $R^{obs}$ for team $t_i$.
For a perfect prediction, the fraction of $\frac{1}{1 + error_i}$ is equal to 1.
MRR values decrease as the error in prediction increases.

While MRR and MAE both consider the deviation values between predicted and observed ranks, MRR imposes higher penalties for higher deviations.
However, similar to previous metrics, it does not distinguish between errors in higher ranks and those in lower ranks.


\subsection{Average Precision}

Average precision (AP) combines list-wise precision and relevance scores to evaluate the performance of a query-response system.
To use AP for evaluating rank prediction in team battle royale, we utilize the extended version of AP introduced in~\cite{dehpanah2020evaluation}.
In a battle royale match with \textit{N} teams, AP can be calculated as:

\small
\begin{equation*}
    AP = \frac{1}{N}\sum_{i=1}^{N} P(i) \times \frac{1}{1 + error_i}
\end{equation*}
\normalsize

\noindent where \textit{P(i)} is the overall precision value up to the \textit{i}\textsuperscript{th} position and $error_i$ is the absolute difference between the predicted rank $R^{pred}$ and observed rank $R^{obs}$ for team $t_i$.

AP uses precision as a weighting factor aiming to reward correct predictions while penalizing incorrect ones.
However, using this component for weighting the predictions may impose an unnecessary strictness over the evaluations since precision focuses on hits and ignores misses even when the difference between predicted and observed ranks is 1 rank.


\begin{figure*}
   \centering
\begin{tabular}{>{\centering\arraybackslash} m{0.7cm} >{\centering\arraybackslash} m{4.3cm} >{\centering\arraybackslash} m{4cm} >{\centering\arraybackslash} m{4cm} >{\centering\arraybackslash} m{4cm} }
\hline
Metric & All Players & Best Players & Frequent Players\\
\hline
\rotatebox[origin=c]{90}{Accuracy}&
\includegraphics[width=4.0cm]{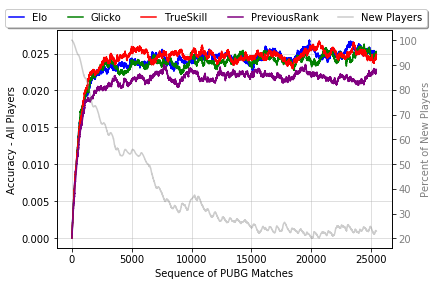}&
\includegraphics[width=3.6cm]{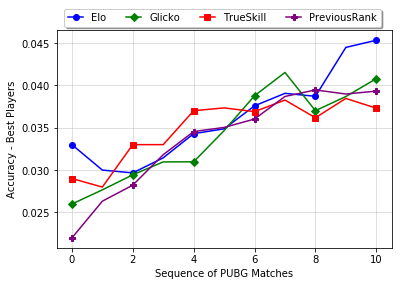}&
\includegraphics[width=3.6cm]{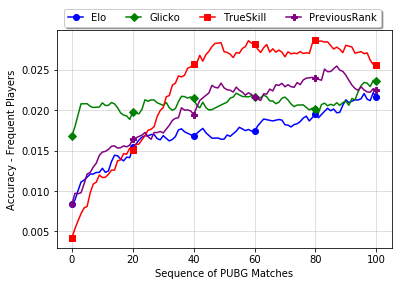}\\
\rotatebox[origin=c]{90}{MAE}&
\includegraphics[width=4.0cm]{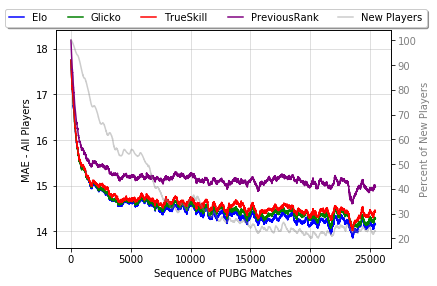}&
\includegraphics[width=3.6cm]{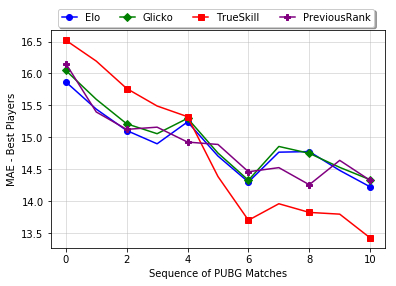}&
\includegraphics[width=3.6cm]{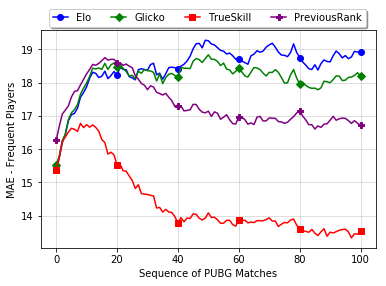}\\
\rotatebox[origin=c]{90}{Tau}&
\includegraphics[width=4.0cm]{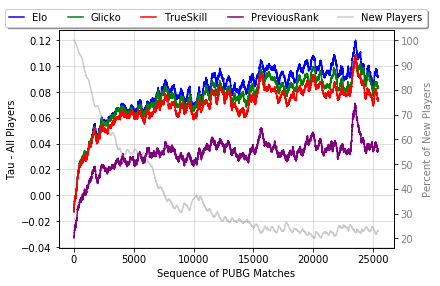}&
\includegraphics[width=3.6cm]{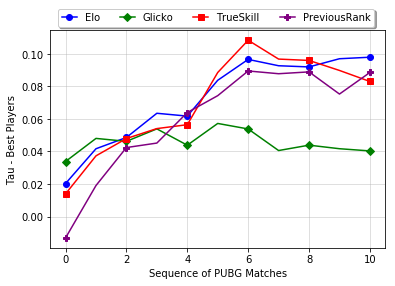}&
\includegraphics[width=3.6cm]{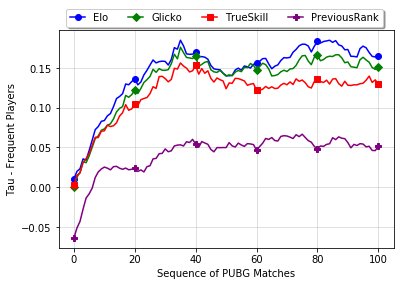}\\
\rotatebox[origin=c]{90}{MRR}&
\includegraphics[width=4.0cm]{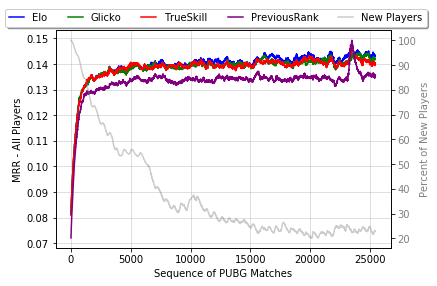}&
\includegraphics[width=3.6cm]{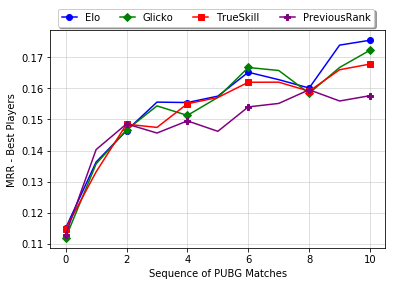}&
\includegraphics[width=3.6cm]{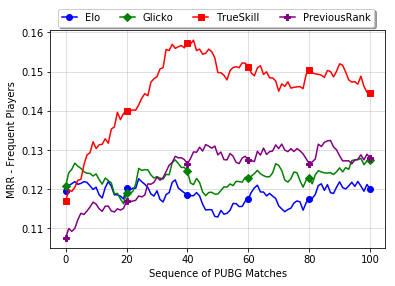}\\
\rotatebox[origin=c]{90}{AP}&
\includegraphics[width=4.0cm]{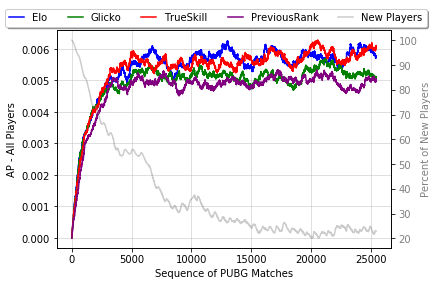}&
\includegraphics[width=3.6cm]{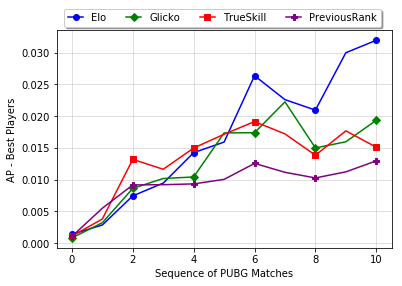}&
\includegraphics[width=3.6cm]{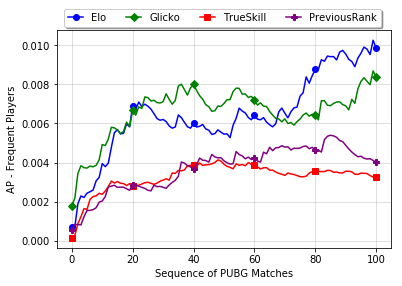}\\
\rotatebox[origin=c]{90}{NDCG}&
\includegraphics[width=4.0cm]{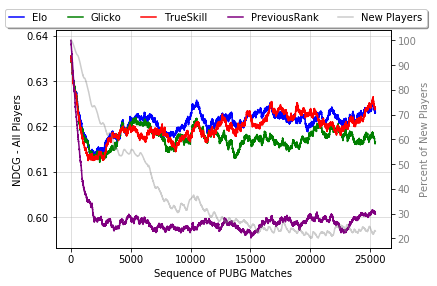}&
\includegraphics[width=3.6cm]{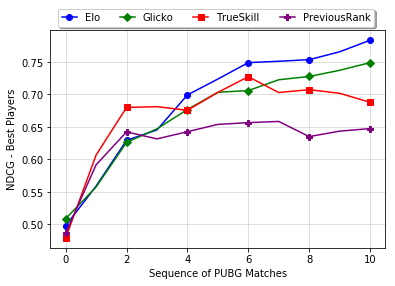}&
\includegraphics[width=3.6cm]{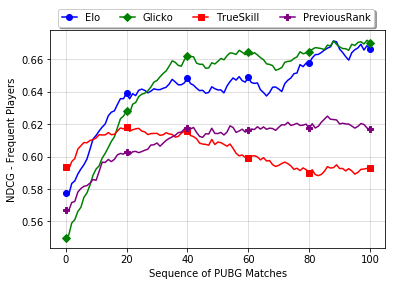}\\
\end{tabular}

\caption{The results of evaluating Elo, Glicko, TrueSkill, and PreviousRank using accuracy, MAE, Kendall's tau, MRR, AP, and NDCG on three different experimental set-ups: all players, best players, and frequent players.}

\label{fig:results} 
\end{figure*}


\subsection{Normalized Discounted Cumulative Gain}
Normalized discounted cumulative gain (NDCG) evaluates the quality of ranked search results based on their relevance scores and positions in the list.
To use NDCG for evaluating rank prediction in team battle royale matches, we utilize an extended version introduced by~\cite{dehpanah2020evaluation}.
In a battle royale match with \textit{N} teams, NDCG is computed as:

\small
\begin{equation*}
    NDCG = \frac{\sum_{i=1}^{N} \frac{1}{log_{2}(i+1)} \times \frac{1}{1+error_i}    }{IDCG}
\end{equation*}
\normalsize

\noindent where $error_i$ is the absolute difference between predicted and observed ranks of team $t_i$. 
$\frac{1}{log(i+1)}$ is the weight assigned to the \textit{i}\textsuperscript{th} position, and \textit{IDCG} is a normalizing factor.

NDCG offers several benefits for evaluating rating systems.
It distinguishes between errors in higher ranks and lower ranks.
It removes the unnecessary strictness of AP by weighting the predictions solely based on their positions.
It also gives the freedom to adjust the weights based on the evaluation goals.
For example, by increasing the weights, the model may direct its attention to evaluating good or regular players who often appear in higher ranks.


\section{Methodology}
\label{sec: methodology}
In this section, we first introduce the dataset used to perform our experiments.
We then detail our methodology.

PlayerUnknown's Battlegrounds (PUBG) is one of the most played games in the battle royale genre.
The game consists of up to one hundred players parachuting onto an island with minimal equipment.
After landing, players start to scavenge for weapons and equipment.
The goal is to eliminate other players and be the last player standing.
PUBG is played in teams of four (squad), teams of two (duo), or singletons (solo).
The dataset is publicly available on Kaggle, a public data platform.

For this research, we only considered duo matches where players compete in teams of two against other teams at the same time.
The filtered dataset provides in-game statistics such as the number of kills, distance walked, and rank for over 25,000 matches and 825,000 unique players.

We first sorted the matches by their timestamps.
For each match, we retrieved the list of teams and players along with their corresponding ratings.
New players were assigned default ratings, 1500 for Elo and Glicko, and 25 for TrueSkill.
For Elo and Glicko, We summed the ratings of team members to calculate the ratings of teams.

We then sorted the teams based on their ratings and used the resulted order as the predicted ranks for the match.
Players' ratings were updated after each match by comparing the predicted ranks and observed ranks of their teams and their corresponding contribution weights.
The parameters we used for each rating system include \textit{k = 10} and \textit{D = 400} for Elo, and $\beta$ = 4.16 and $\tau$ = 0.833 for TrueSkill as suggested in its official documentation. 

We evaluated the performance of rating systems using three different set-ups.
In the first set-up, we considered all players and all unique matches in the dataset sorted by date.
In this set-up, every player is treated equally regardless of their skill or how often they play the game.
This set-up includes all of the new players in the system and thus, provides the opportunity to investigate how the explanatory power of metrics is impacted by introducing new players in each match.

Second, we evaluated the performance of the rating systems for the best players in the system.
To identify the best players, we sorted all players by their most recent rating and selected the top 1000 players who had played more than 10 games.
Since these players did not compete at the same time, we evaluated the predictive performance of the rating systems on their first 10 games.
Focusing on the best players provides the opportunity to study how metrics evaluate the ability of the systems on capturing the skill level of players who presumably show consistent playing behavior and often achieve the highest ranks in the games they play.

Finally, we evaluated the performance of rating systems for the most frequent players in the system.
To identify these players, we selected all players who played more than 100 games.
We evaluated the predictive performance of the rating systems on their first 100 games.
In this scenario, the main assumption is that playing more games leads to higher consistency in playing behavior and higher skills, and as a result, rating systems can achieve more accurate skill estimations as they observe more games from players.
Focusing on the most frequent players provides an additional way to study how metrics evaluate the ability of rating systems on learning more about the players by observing more games from them.


\section{Results and Discussions}
\label{sec: results}

In this section, we present and discuss the results of three experiments on all players, best players, and most frequent players.
We compare the performance of the competing models using six metrics including accuracy, MAE, Kendall's tau, MRR, AP, and NDCG.
We analyze the ability of these metrics in explaining the predictive behavior of rating systems.

The results of the experiments are given in Figure~\ref{fig:results}.
In this figure, rows correspond to evaluation metrics and columns are associated with experimental set-ups.
The performance of the models is represented by trend lines with different colors.


\subsection{Accuracy}

Accuracy was extensively used for evaluating rank predictions in head-to-head matches where there are only two possible outcomes (three, if a draw is possible).
However, as shown in figure \ref{fig:results}, the resulted accuracy values in all three set-ups are very small compared to those often observed for head-to-head matches.
We expect the rating systems to gain a better knowledge of players and improve their skill estimations by observing more games.
This improvement may depend on several factors such as the consistency in playing behavior of players and their frequency of play.

For the `all players' set-up, accuracy trends indicate that rating systems could successfully achieve more accurate predictions over time.
However, accuracy values are much smaller at the start of the sequence of matches where most of the players are newly added to the system with default rating values.
These matches contain all or many ties in ratings that are randomly broken by rating systems.
The rating systems show increasing accuracy as the number of new players decreases (shown by the gray trend line).
This trend continues at a fast rate up to the point where the percentage of new players in matches reaches less than 60\% after which accuracy values remain almost fixed around 2.5\%.
While Elo, Glicko, and TrueSkill outperform PreviousRank in this scenario, no distinct advantage can be observed among themselves; all three achieve around 2.5\% accuracy at the end of the sequence where almost 80\% of the players are known to the system.

The next set-up, `best players', considers players who often show consistent playing behavior and achieve the best ranks in matches they play.
Therefore, we assume the rating systems estimate their true skill level faster and more accurately.
The accuracy patterns observed for this scenario confirms this assumption.
Accuracy values are almost twice as much for the best players compared to those for the `all players' set-up.
However, it seems that accuracy cannot distinguish between the performance of the rating systems and our naive baseline for this set-up.
All competing models demonstrate increasing accuracy values over the course of 10 games.
This finding suggests that, regardless of the skill level of their teammates, skilled players could achieve higher ranks in matches they played either by carrying their teammates or by ignoring the team and playing on their own.

For the next set-up, we considered players who compete more frequently than others.
We assume that playing more games improves players' skills and consequently results in more accurate skill estimations by reducing the uncertainties about their performance.
The accuracy patterns observed in this scenario confirm this assumption.
The patterns are almost similar to those for the `all players' set-up except that the accuracy values start from 0.5\% instead of zero.
The rating systems could increase their knowledge of frequent players over the course of 100 games.
This finding shows that, even when teaming up with new or inexperienced players, frequent players could show relatively consistent playing behavior and lead the team effort to guide their teammates by communicating through voice, text, or radio messages.

Accuracy could capture the predictive behavior of rating systems to some extent.
However, as small accuracy values observed in all three set-ups suggest, it fails to capture their real performance in battle royale matches because when the number of possible outcomes increases, there is a smaller chance of scoring hits in rank predictions.
To achieve a better understanding of the real predictive power of rating systems, it is better to use metrics that do not treat ranks as labels.


\subsection{MAE (Mean Absolute Error)}

Compared to accuracy, MAE is more suited for evaluating numerical values as it considers the deviations between values to compare their similarity. 
As depicted in Figure~\ref{fig:results}, the results of evaluating rating systems using MAE show significant errors in rank predictions.

For the `all players' set-up, decreasing MAE trends confirm our assumption that rating systems achieve more accurate rank predictions by observing more games from players over time.
As mentioned before, the early stages of the sequence of matches mainly consist of new players about whom the rating systems do not possess any knowledge.
That is why the error values are much higher for these matches.
Rank prediction errors decrease as the number of new players decreases over the sequence.
As the percentage of new players in matches reaches less than 40\%, the decreasing rate of MAE becomes slower to a point where MAE values remain almost fixed around 14.5.
Elo, Glicko, and TrueSkill outperformed PreviousRank in this set-up with Elo slightly outperforming the other two rating systems.

For the next set-up, we considered the best players representing players with consistent playing behavior in the system.
We expect the rating systems to generate more accurate rank predictions for these players and capture their true skill level faster.
Based on the MAE patterns observed in this scenario, rating systems partly meet our expectation.
Compared to the MAE of 18 for the early stages of the `all players' set-up, MAE values here start from around 16.
However, Elo and Glicko show the same performance level for the `all players' and `best players' set-ups at the end of the sequence, finishing with MAE of around 14.5.
On the other hand, MAE values for TrueSkill start from 16.5 and reach 13.5 at the end of the sequence, showing a small improvement compared to those of the `all players' set-up.
However, MAE is unable to draw a distinction between the performance of the rating systems and PreviousRank for this set-up.

For the `frequent players' set-up, the underlying assumption is that playing more games leads to more consistent playing behavior from players, and observing more games from such players helps the rating systems to achieve more accurate skill estimations and rank predictions.
MAE patterns for the `frequent players' set-up suggest different behaviors from rating systems.
All models show increasing MAE values at the start of the sequence.
Among the three rating systems, only TrueSkill could correct the trend and reduce prediction errors.
While the `frequent players' set-up removes the influence of new players on rank predictions, Elo and Glicko show higher errors in this set-up compared to predictions in the `all players' set-up that suffer from this influence.
Based on MAE values for this scenario, our assumption cannot be confirmed.

MAE only considers the deviations between predicted and observed ranks regardless of rank positions and thus, fails to capture the evaluation details.
Metrics that consider rank positions could provide more accurate evaluations.


\subsection{Kendall's tau}
Unlike accuracy and MAE, Kendall's tau considers rank positions when evaluating the similarity between two sets of ordered values.
The results of evaluating the predictive performance of rating systems using Kendall's tau show fairly small correlations between predicted and observed ranks.

For the `all players' set-up, negative tau values observed at the early stages of the sequence suggest that its evaluations are highly interrupted by the influence of new players to a degree that predicted ranks and observed ranks have the opposite order.
However, all models show increasing tau values as the number of new players in the system decreases suggesting their ability to improve skill estimations by observing more games and learning more about the behavior of players.
Elo, Glicko, and TrueSkill significantly outperform PreviousRank showing fairly similar performance levels.

For the `best players' set-up, our assumption is that rating systems generate more accurate predictions for the best players and learn their true skill level faster since they reduce the systems' uncertainties by showing more consistent playing behavior.
Tau patterns in this scenario show that it could capture the learning ability of Elo and TrueSkill but fails to do so for Glicko.
Also, while this set-up removes the interrupting influence of new players, tau values are about the same as the values observed in the previous set-up where that influence was obvious.
On the other hand, compared to the `all players' set-up, tau values for our non-parametric baseline, PreviousRank, show higher improvements than those for the rating systems.
In addition, tau is unable to distinguish between the performance of the rating systems and PreviousRank for this set-up.
These findings suggest that Kendall's tau fails to correctly capture the predictive performance of rating systems for the best players.

The next set-up, `frequent players', suggests the opposite.
The increasing tau trends for all rating systems and higher values compared to the two previous set-ups in this scenario indicate that tau captures the learning ability of rating systems by observing more games from the players.
Elo, Glicko, and TrueSkill outperform PreviousRank by a large margin.

Our results suggest that Kendall's tau is not an appropriate metric for evaluating rank predictions in team battle royale matches.
Tau evaluations are highly interrupted by the influence of new players.
It also cannot correctly capture the learning patterns of rating systems for top-tier players.
Finally, tau values observed in all set-ups suggest that it cannot represent the real predictive power of rating systems.
Kendall's tau only considers the pairwise agreement between predicted and observed ranks without regard to their deviations.
Therefore, it cannot capture the real differences between players or teams.


\subsection{MRR (Mean Reciprocal Rank)}

MRR averages the relevance scores of predictions to evaluate the predictive performance of the models.
The evaluation patterns observed for MRR is partly similar to those for the previous metrics.

As seen in Figure~\ref{fig:results}, The results of the first set-up, `all players', show increasing MRR trends for all models over the sequence of matches.
These trends suggest that MRR is capable of capturing the learning ability of rating systems from observing more games.
On the other hand, MRR evaluations are clearly interrupted by the influence of new players.
However, when the percentage of new players in matches reaches less than 70\%, MRR trends become almost steady remaining around 14\%.
Elo, Glicko, and TrueSkill demonstrate slightly higher MRR values than PreviousRank but MRR values are almost similar for all three rating systems.

MRR patterns for the next set-up, `best players', show clear evidence of better performance and learning ability of rating systems for players who are assumed to show consistent playing behavior.
The MRR values are higher in this scenario compared to the previous set-up while also showing an almost constant increase over the course of 10 games.
MRR is also able to slightly capture the superiority of rating systems over our naive baseline for this scenario.
Elo, Glicko, and TrueSkill show almost similar performance over the whole sequence.

The MRR evaluation results for frequent players suggest the opposite.
There is no clear improvement observed for Elo and Glicko over the course of 100 games.
TrueSkill shows a considerable improvement at the start, achieving MRR of around 16\% after the 40\textsuperscript{th} game, but fails to maintain it and gradually drops down and reaches MRR of around 14.5\% at the end of the sequence.
In addition, unlike the previous set-up, in this scenario, PreviousRank outperforms both Elo and Glicko.
Finally, since the `frequent players' set-up removes the influence of new players, we expect the rating systems to perform better in this set-up than the `all players' set-up.
However, while MRR values for TrueSkill are almost the same in both set-ups, Elo and Glicko demonstrate smaller MRR values in this set-up.
These findings show that MRR is not capable of capturing the true predictive behavior of rating systems based on the frequency of play.

While MRR performed well for the `all players' and `best players' set-ups, the patterns observed in the `frequent players' set-up question its ability for evaluating battle royale matches.
MRR offers generic rank-based evaluations without distinguishing between prediction errors in higher ranks and lower ranks.
Using metrics that adjust their scores based on the prediction error's position may benefit the evaluations.


\subsection{AP (Average Precision)}

AP leverages precision values up to each rank to weight the predictions based on their position.
The results of evaluating rating systems using AP suggest that although AP values are significantly low, it is able to capture the true predictive behavior of rating systems.

The AP patterns for the `all players' set-up is fairly similar to what we observed for accuracy in the same set-up.
AP values start from zero and continue increasing at a fast rate as the number of new players in the system decreases.
This rate starts to slow down when the percentage of new players in matches drops below 60\% after which AP values remain almost fixed around 0.55\%.
The observed values are extremely small because AP uses a hit-based scoring function-- precision-- for weighting the predictions.
While previous metrics could not draw a clear distinction between the performance of the rating systems in the `all players' set-up, AP seems to be able to slightly distinguish between their performance in this scenario.
However, AP cannot make a clear distinction between the performance of PreviousRank and Glicko in this set-up.

The AP values increased in the `best players' set-up showing that it could capture the learning ability of the rating systems from observing more games from players with consistent playing behavior.
The rating systems demonstrate better predictive performance compared to PreviousRank.

For the `frequent players' set-up, all models show increasing AP trends.
The AP values observed for Elo and Glicko in this scenario are significantly smaller than those in the `best players' set-up, but are slightly higher than those in the `all players' set-up.
This means that AP correctly evaluated their predictive behavior.
However, it seems that AP cannot capture the true predictive power of TrueSkill as not only was it outperformed by PreviousRank, but it also achieved slightly smaller values compared to the `all players' set-up; the set-up that suffers from the influence of new players.

AP is able to detect the difference between players or teams as it considers both deviations and positions when comparing predicted ranks and observed ranks.
However, due to using precision values as the weighting factor, it highly underestimates the true predictive power of rating systems as the extremely small values observed suggest.
Using a non-binary weighting factor that looks beyond scoring hits may alleviate this issue.


\subsection{NDCG (Normalized Discounted Cumulative Gain)}

NDCG leverages a non-binary factor to weight predicted ranks based on the position of the observed ranks.
As the patterns suggest, NDCG can correctly capture both predictive power and behavior of rating systems in most cases.

For the `all players' set-up, NDCG evaluations are interrupted by the influence of new players when at least 80\% of the players in the match are new to the system.
NDCG values slowly increase as the number of new players in the system decreases over the sequence of matches.
The patterns suggest that NDCG delivers the best evaluations among all metrics in terms of capturing the superiority of the rating systems over our naive baseline.
It also draws a slightly more clear distinction between the performance of rating systems compared to the other metrics.

For the `best players' set-up, rating systems achieve much higher NDCG values showing their ability to generate more accurate skill estimations and rank predictions by observing more games from players with consistent playing behavior.
NDCG patterns indicate that rating systems significantly outperform PreviousRank.
The distinction among the performance of rating systems is also clear in this scenario. 

The results of the `frequent players' set-up are fairly close to those observed for AP.
Compared to the `all players' set-up, the NDCG values in this scenario are significantly higher for Elo and Glicko.
TrueSkill shows a decreasing trend to a point that it is even outperformed by our naive baseline-- PreviousRank.
These findings suggest that NDCG correctly captures the predictive power and behavior of Elo and Glicko for the most frequent players but fails to do so for TrueSkill.

NDCG alleviates most of the challenges faced by other metrics.
Similar to AP, NDCG considers both rank positions and deviations when comparing predicted ranks with observed ranks.
However, instead of binary weights, NDCG calculates prediction scores based on the positions using a rank-based weighting factor.
This way, it distinguishes between the errors in higher ranks with those in lower ranks.
Finally, based on all the patterns observed, NDCG seems to be the best metric for evaluating rank predictions in team battle royale games.


\section{Conclusion and Future Work}
\label{sec: conclusion}

In this paper, we used several metrics to evaluate the predictive performance of three popular rating systems in team battle royale matches.
We performed our experiments on three different set-ups to evaluate different aspects of rank predictions.
We first evaluated the predictive power of rating systems considering all players, teams, and matches in the dataset to investigate how the systems perform over time under the influence of new players about whom the systems do not possess any knowledge.
We also evaluated rank predictions for the best players in the dataset to investigate the influence of playing behavior on the evaluations.
Finally, we evaluated the predictive power of rating systems on the most frequent players in the dataset to investigate the influence of the frequency of play on the evaluations.

Our evaluation results show stark differences in the utility of tested metrics.
Metrics were affected by the influence of new players at different rates.
For example, NDCG is able to overcome this influence and show reliability when less than 80\% of players in the match are new to the system while this value for MAE is around 55\%.
Some metrics could capture the ability of rating systems to achieve a better knowledge of players by observing more games.
Others demonstrated different patterns for each set-up.
Some metrics were also able to draw clear distinctions between the performance of the rating systems.

Among all metrics tested, NDCG provided more reliable evaluations while capturing both predictive power and behavior of rating systems.
It resolved most of the challenges faced by other metrics.
Compared to other metrics, NDCG made the clearest distinction between the performance of the rating systems while showing their superiority to PreviousRank.

This work was part of our effort to achieve a better understanding of the evaluation of rating systems and their rank prediction.
We plan to improve the predictive power of rating systems by incorporating players’ behavioral features as their input and modeling players and teams based on interactions occurring within the team and between the competing teams in a match.
We will then extend rank prediction to building a framework for predicting the success of proposed teams and making assignments.


\bibliographystyle{IEEEtran}
\bibliography{IEEEabrv, main}

\end{document}